\documentclass[12pt]{article}

\usepackage{graphicx}

\begin{document}

\begin{centering}
{\Large \bf Lunar Laser Ranging, Gravitomagnetism and Frame-Dragging}\\
\vspace{.25in}

Ignazio Ciufolini\\

\vspace{.25in}

Dipartimento di Ingegneria dell'Innovazione, Universit\`{a} di
Lecce and INFN sezione di Lecce, Via Monteroni, 73100 Lecce, Italy\\
\vspace{.9in}
\end{centering}

\centerline {\large \bf Abstract}

{During the past century Einstein's theory of General Relativity
gave rise to an experimental triumph, however, there are still
aspects of this theory to be measured or more accurately tested. One
of the main challenges in experimental gravitation, together with
the direct detection of gravitational waves, is today the accurate
measurement of the gravitomagnetic field generated by the angular
momentum of a body. Here, after a description of frame-dragging and
gravitomagnetism and of the main experiments to detect these
relativistic phenomena, we show that the fundamental tests of
General Relativity performed by Lunar Laser Ranging do not, however,
include a measurement of the intrinsic gravitomagnetic field
generated by the angular momentum of a body}.

\vspace{.25in}

\noindent {\it Dedicated to John Archibald Wheeler, a master of
physics of the XX century and father of the renaissance of General
Relativity}

\newpage

\section{Introduction}

A number of experiments have been proposed and performed to
accurately measure the gravitomagnetic field generated by the
angular momentum of a body and frame-dragging \cite{ciu07,tho,ciuw},
from the complex space experiment Gravity Probe B, launched by NASA
in 2004 after more than 40 years of preparation \cite{gpb}, to the
observations of the LAGEOS and LAGEOS 2 satellites
\cite{ciu04,ciu06} and from the LARES satellite, to be launched in
2009 by ASI (Italian Space Agency) \cite{ciu06} using the new
launching vehicle VEGA of ESA (European Space Agency), to Lunar
Laser Ranging \cite{llr}, binary pulsars \cite{sta} and other
astrophysical observations \cite{nord,cui}, including a number of
other space experiments currently proposed to international space
agencies.

In Einstein's gravitational theory the local inertial frames have a
key role \cite{mtw,wei,ciuw}. The strong equivalence principle, at
the foundations of General Relativity, states that the gravitational
field is locally 'unobservable' in the freely falling frames and
thus, in these local inertial frames, all the laws of physics are
the laws of Special Relativity. The local inertial frames are
determined, influenced and dragged by the distribution and flow of
mass-energy in the Universe; the axes of these non-rotating, local,
inertial frames are determined by torque-free test-gyroscopes that
are dragged by the motion and rotation of nearby matter, for this
reason this phenomenon is called dragging of inertial frames or
frame-dragging \cite{ciuw,ciu07}.

In General Relativity, a torque-free spinning gyroscope defines an
axis non-rotating relative to the local inertial frames; the orbital
plane of a test particle is also a kind of gyroscope. The
frame-dragging effect on the orbit of a satellite, due to the
angular momentum vector $\vec J$ of a central body, is known as
Lense-Thirring effect: ${\vec \Omega}_{L-T} = {{2 G \vec {J}} \over
{c^2 a^3 (1-e^2)^{3/2}}}$, where ${\vec {\Omega}}_{L-T}$ is the rate
of change of the longitude of the nodal line of the satellite, that
is the intersection of its orbital plane with the equatorial plane
of the central body, i.e., it represents the rate of change of the
orbital angular momentum vector, $a$ is the semi-major axis of the
orbiting test-particle, $e$ its orbital eccentricity, $G$ the
gravitational constant and $c$ the speed of light. The
frame-dragging by the Earth spin has been measured using the LAGEOS
satellites with an accuracy of the order of 10 percent
\cite{ciu04,ciu06}, might be detected by further Gravity Probe B
data analysis \cite{gpb} and will be measured with improved accuracy
by the LARES satellite.

\section{Lunar Laser Ranging, gravitomagnetism
and geodetic precession}

In General Relativity there is another type of frame-dragging effect
and precession of a gyroscope known as geodetic precession or de
Sitter effect \cite{ciuw,ciu07}. If a gyroscope is at rest with
respect to a non-rotating mass, it does not experience any drag.
However, if the gyroscope starts to move with respect to the
non-rotating mass it acquires a precession that will again disappear
when the gyroscope will stop relative to the non-rotating mass. The
geodetic precession, due to the velocity $\vec v$ of a
test-gyroscope, is: ${\vec \Omega}_{geodetic} = {3 \over 2} {G M
\over {c^2 r^3}} \vec x \times  \vec v$, where $M$ is the mass of
the central body and $\vec x$ and $r$ are position vector and radial
distance of the gyroscope from the central mass.

A basic difference between frame-dragging by spin and geodetic
precession is that in the case of the former (the Lense-Thirring
effect) the frame-dragging effect is due to the additional spacetime
curvature produced by the rotation of a mass, whereas in the case of
the latter (the de Sitter effect) the frame-dragging effect is due
to the motion of a test-gyroscope on a static background and its
motion produces no spacetime curvature, (see below and section 6.11
of ref. \cite{ciuw}; for a discussion on frame-dragging and geodetic
precession see refs \cite{ashs,ocon,ciu07}).

The geodetic precession has been measured on the Moon's orbit by LLR
with accuracy of the order of 0.6 percent \cite{ciu87,llr,llr2}, by
Gravity Probe B with approximately 1 percent accuracy \cite{gpb} and
has been detected on binary pulsars \cite{tay,sta}.

Lunar Laser Ranging (LLR) is a basic tool for testing fundamental
physics and General Relativity. By short laser pulses, the range
from an emitting laser on Earth and a retro-reflector on the Moon is
today measured with an accuracy of the order of a centimeter,
corresponding to a fractional error in the distance of approximately
$2.6 \times 10^{-11}$. In addition to the important applications of
LLR for the study of the dynamics of the Earth-Moon system and of
the Moon internal structure, in fundamental physics LLR has provided
accurate tests of the strong and weak equivalence principle,
accurate measurements of the PPN (Parametrized Post Newtonian)
parameters testing General Relativity \cite{wil}, experimental
limits on conceivable time variations of the gravitational constant
G and accurate tests of the geodetic precession \cite{llr,llr2}.

Recently, a number of authors have debated whether the
gravitomagnetic interaction and frame-dragging by spin have also
been accurately measured on the Moon orbit by Lunar Laser Ranging
\cite{mnt,kop,mnt2,ciuxxx}. This is a recent chapter of a long
debate on the meaning of frame-dragging and gravitomagnetism
\cite{ashs,ocon,ocon2,ocon3,mnt,kop,mnt2,ciuxxx,ciu1,ciuw}; a basic
issue treated in \cite{mnt,kop,mnt2} is whether the effect detected
by LLR is a frame-dependent effect or not.

In order to answer to this question, we propose here a distinction
between gravitomagnetic effects generated by the translational
motion of the frame of reference where they are observed, e.g., by
the motion of a test-gyroscope with respect to a central mass (not
necessarily rotating), and those generated by the rotation of a mass
or by the motion of two masses (not test-particles) with respect to
each other, without any necessary motion of the frame of reference
where they are observed. The geodetic precession is a translational
effect due to the motion of the 'Earth-Moon gyroscope' in the static
field of the Sun. The Lense-Thirring effect measured by the LAGEOS
satellites, that might also be detected by further Gravity Probe B
data analysis and by LARES, is due to the rotation of a mass, i.e.,
by the rotation of the Earth mass. In the following we show that the
gravitomagnetic effect discussed in \cite{mnt} is just a
translational effect that is substantially equivalent to the Moon's
geodetic precession. For this purpose, a rather illuminating formal
analogy of General Relativity with electrodynamics is briefly
described in the next section.

\section{Gravitomagnetism and Electromagnetism}

Whereas in electrodynamics an electric charge generates an electric
field and a current of electric charge produces a magnetic field, in
Newtonian gravitational theory the mass of a body generates a
gravitational field but a current of mass, for example the rotation
of a body, does not produce any additional gravitational field. On
the other hand, Einstein's gravitational theory predicts that a
current of mass generates an additional gravitomagnetic field that
exerts a force on surrounding bodies and changes the spacetime
structure by generating additional curvature.

In General Relativity, the gravitomagnetic field due to the angular
momentum ${\vec {J}}$ of a central body is, in the weak-field and
slow-motion approximation:

\begin{equation}
{\vec {H}} ~ = ~ {\vec {\nabla}}~ \times ~ {\vec {h}} ~ \cong ~ 2 ~
G \Biggl [{{\vec {J}} \,- \,3 ({\vec {J}} ~ \cdot ~ {\hat x})~ {\hat
{x}} \over {c^3 r^3}} \Biggr ]
\end{equation}

\noindent where $r$ is the radial distance from the central body,
$\hat x$ is the position unit-vector and $\vec h$ is the so-called
'gravitomagnetic vector potential' (equal to the non-diagonal, space
and time, part of the metric). The gravitomagnetic field generates
frame-dragging of a gyroscope in a way formally similar to the
magnetic field producing a change of orientation of a magnetic
needle (magnetic dipole). Indeed, in General Relativity, a current
of mass in a loop, that is a gyroscope, has a behavior formally
similar to that of a magnetic dipole in electrodynamics which is
made of an electric current in a loop (see Fig. 1). The precession
${\vec {\Omega}}_S$ of the spin axis of a test-gyroscope by the
angular momentum $\vec {J}$ of a central body is: ${\vec {\Omega}_S}
~ = ~ {{3 \, ( G {\vec {J}} ~ \cdot ~ {\hat {x}}) ~ \hat { {x}} - {G
\vec  {J}}} \over {c^2 r^3}}$, where ${\hat x}$ is the position
unit-vector of the test-gyroscope and $r$ its radial distance from
the central body.

In electromagnetism, in a frame where a test-particle with electric
charge is at rest we only observe an electric field $E^k$ but no
magnetic field, however, in a frame that is moving relative to the
charge we also measure a magnetic field $B^k$. In General
Relativity, in a similar way, in a frame where a non-rotating mass
is at rest, the components of the gravitomagnetic vector potential
$h^k$ are zero. Nevertheless, if we consider an observer moving
relative to the mass, in a local frame moving with the observer the
components of the gravitomagnetic vector potential $h^k$ are
non-zero but can of course be annulled by a Lorentz transformation
back to the original frame. Indeed, in a frame where a non-rotating
mass {\it M} is at rest, the only components of the Schwarzschild
metric $g_{\alpha \beta}$ different from zero (written in standard
Schwarzschild coordinates) are: $g_{00} = - g^{-1}_{rr} = - \left(1
- {2 G M \over {c^2 r}} \right)$, $g_{\theta \theta} = r^ {2}$ and
$g_{\phi \phi} = r^ {2} sin^{2} \theta$ and the three non-diagonal
components of the metric $g_{0k}$, i.e., the components of the
'gravitomagnetic vector potential' $h^k$, are zero. Nevertheless, if
we perform a local Lorentz transformation with velocity $v^k$
relative to the mass {\it M}, the components of the gravitomagnetic
vector potential $g_{0k}$ are non-zero in the new frame. The orbital
effects of this gravitomagnetic vector potential, arising from
motion of the Earth-Moon system relative to the Sun mass, have been
observed by LLR since the first measurements of the geodetic
precession of the Moon orbit, i.e., of the Earth-Moon 'gyroscope'
moving around the Sun. On the other hand, the angular momentum $\vec
J$ of a body generates a gravitomagnetic field and produces
spacetime curvature that cannot be eliminated by a simple change of
frame of reference or by a coordinate transformation. This
gravitomagnetic field generates the Lense-Thirring effect on the
orbit of the LAGEOS satellites.

In order to distinguish between 'intrinsic' gravitomagnetic effects
(the Lense-Thirring effect) and 'translational' ones (the geodetic
precession), we have proposed to use spacetime curvature invariants.
Here, below, we show that the phenomenon discussed in
\cite{mnt,mnt2} is a translational gravitomagnetic effect. In
general, one cannot derive intrinsic gravitomagnetic effects from
translational ones unless making additional theoretical hypotheses,
such as the linear superposition of the translational
gravitomagnetic effects; for example, the magnetic field generated
by the intrinsic magnetic moment (Bohr magneton) is an intrinsic
phenomenon due to the intrinsic spin of a particle that cannot be
explained and derived as a translational effect by any Lorentz and
frame transformation.

In electromagnetism, in order to characterize the electromagnetic
field, using the electromagnetic field Lorentz-tensor $F_{\alpha
\beta}$ we can build the scalar Lorentz-invariant ${\bf ^\ast F
\cdot F} \equiv {1 \over 4} F_{\alpha \beta} \; ^\ast F^{\alpha
\beta} = {\vec E\/} \cdot {\vec B \/}$, where $^\ast F^{\alpha
\beta}$ is the dual of $F^{\alpha \beta}$, defined as: $^\ast
F^{\alpha \beta} = {1 \over 2} \varepsilon^{\alpha \beta \mu \nu} \,
F_{\mu \nu}$ and $\varepsilon^{\alpha \beta \mu \nu}$ is the
Levi-Civita pseudotensor (that is equal to $+ {\sqrt {-g}}$, i.e.,
plus the square root of minus the determinant, $g$, of the metric,
if the indices are even permutations of (0,1,2,3), $- {\sqrt {-g}}$
for odd permutations of (0,1,2,3) and 0 if any indices are
repeated). ${\bf ^\ast F \cdot F}$ is an invariant for Lorentz
transformations (precisely a pseudo-invariant under coordinate
reflections), i.e., is either null or not in every inertial frame.
For example, in the rest frame of a test-particle with charge $q$ we
have an electric field only and no magnetic field, and this
invariant is zero, therefore even in a frames moving relative to
$q$, where both ${\vec B} \not = {\vec 0\/}$ and ${\vec E \/} \not =
{\vec 0 \/}$, this invariant is zero. However, in a frame where a
charge $q$ and a magnetic dipole ${\vec m \/}$ are at rest, we have
in general ${\bf ^\ast F \cdot F} \not = 0 $ and therefore this
invariant is non-zero in any other inertial frame.

In General Relativity, the gravitomagnetic 'vector' potential $h^k$
can be zero or not depending on the frame where it is calculated.
Nevertheless, the curvature of a manifold is a coordinate
independent quantity \cite{mtw,wei,ciuw}. Therefore, in order to
test for intrinsic gravitomagnetic effects, i.e., independent of the
coordinate system (and not eliminable with a coordinate
transformation) we have to use the Riemann curvature tensor
$R_{\alpha \beta \mu \nu}$ and the spacetime invariants built with
it \cite{ciu1,ciuw}. Given a metric $g_{\alpha \beta}$ in some
coordinate system (with or without the so-called 'magnetic'
components $g_{0k}$), in a way similar to electromagnetism, using
the Riemann curvature tensor $R_{\alpha \beta \mu \nu}$ we can build
the spacetime curvature invariant ${\bf {^\ast R \cdot R}} \equiv
{^\ast R}^{\alpha \beta \mu \nu} \; R_{\alpha \beta \mu \nu}$, where
${^\ast R}^{\alpha \beta \mu \nu} \equiv {1 \over 2} \;
\varepsilon^{\alpha \beta \sigma \rho} R_{\sigma \rho} ~ ^{\mu \nu}$
is the dual of $R_{\alpha \beta \mu \nu}$ \cite{ciuw}.   Here below
and in \cite{ciuw} the exact explicit expression of the Riemann
curvature invariant $\bf ^\ast R \cdot R$ is given for some
spacetime solutions of the Einstein field equation. For example, in
the case of the Kerr metric generated by the angular momentum $J$
and the mass $M$ of a rotating body, this invariant (precisely a
pseudo-invariant under coordinate reflections) is equal to
\cite{math}:

\begin{equation}
{1\over 2} \varepsilon ^{\alpha \beta \sigma \rho } R _ {\sigma
\rho} ~ ^{\mu \nu} R_ {\alpha \beta \mu \nu } = 1536 \, J \, M ~cos
\theta { \left( r^5 \rho ^{-6} - r^3 \rho^{-5} +{ 3\over 16} r \rho^
{-4} \right) } \;\;\;
\end{equation}

where $\rho = \left ({r^2 + \Bigl ( {J \over M} \Bigr )^2 cos^2
\theta} \right)$; this expression of ${\bf {^\ast R \cdot R}}$ is
then different from zero if and only if $J \not = 0$, e.g., it is
zero in the case of the Schwarzschild metric generated by the mass
only of a non-rotating body (with $J = 0$). In the case of Earth
with angular momentum $J_{\oplus}$, the invariant $\bf {^\ast R
\cdot R}$ is at the lowest order: $^\ast {\bf R} \cdot {\bf R}
\simeq 288 ~ {G^2 J_\oplus M_\oplus \over {c^5} r^7} cos \theta + .
. . $, where $\theta$ is the colatitude, thus the Lense-Thirring
effect on the LAGEOS satellites is an intrinsic gravitomagnetic
effect \cite{ciu1,ciuw} that cannot be eliminated by a change of
frame of reference. However, in the next section we show that the
effect discussed in \cite{nor03,mnt}, accurately measured by Lunar
Laser Ranging, is just a 'translational' gravitomagnetic effect
which depends on the frame of reference used in the analysis; the
invariant $\bf {^\ast R \cdot R}$ is indeed null on the ecliptic
plane (apart from the intrinsic gravitomagnetic terms due to
$J_\oplus$ and $J_\odot$) and the gravitomagnetic term discussed in
\cite{nor03,mnt}, when analyzed in a different frame, is
substantially equivalent to the geodetic precession.

\section {\large Lunar Laser Ranging and gravitomagnetic effects}

In \cite{nor03,mnt} is analyzed a gravitomagnetic perturbation of
the Moon orbit consisting in a change of the Earth Moon distance of
about 5 meters with monthly and semi-monthly periods. This variation
of the Earth Moon distance is, in the Moon's geodesic equation of
motion, due to the gravitomagnetic acceleration \cite{nor03}:
\begin{equation}
{\vec a_I} = {4 \over c^2} \sum_{J \neq I} \vec{v}_I \times
(\vec{v}_J \times \vec{G}_{IJ}),
\end{equation}
where the index $I$ indicates the Moon and $J$ the Sun and Earth,
$\vec{v}_I$ and $\vec{v}_J$ are their velocities, $\vec{G}_{IJ} = {G
M_J \over r^3_{IJ}} \vec{r}_{IJ}$ is the standard Newtonian
acceleration vector, $\vec{r}_{IJ}$ the position vector from body
$I$ to body $J$ and $\vec{r}_{IJ} = | \vec{r}_{IJ} |$.

In a frame of reference comoving with the Sun, we find the
gravitomagnetic acceleration (2) of the Moon:

\begin{equation}
{4 \over c^2} \; \vec{v}^{(S)}_{M} \times (\vec{v}^{(S)}_\oplus
\times \vec{G}_{M \oplus})
\end{equation}

\noindent i.e, the term analyzed in \cite{nor03,mnt}, where the
upper letter within parenthesis indicates if the corresponding
quantity is measured with respect to the Sun: $^{(S)}$, or to Earth:
$^{(E)}$, i.e., $\vec{v}^{(E)}_{M}$ is the velocity of the Moon with
respect to Earth and $\vec{v}^{(S)}_\oplus$ is the velocity of Earth
with respect to the Sun. In the Moon's equation of motion there is
another gravitomagnetic term due to the velocity of the Earth-Moon
system around the Sun:

\begin{equation}
2 \, {\vec \Omega}_{geodetic} \times \vec{v}^{(E)}_{M}
\end{equation}

where $\vec{\Omega}_{geodetic} = {3 \over 2} {G M_\odot \over {c^2
R^3}} \vec{R} \times  \vec{v}^{(S)}_\oplus$ is the geodetic
precession of a gyroscope comoving with Earth and $R$ is the radial
distance of Earth from the Sun.

The discussion of the effect of term (2) on the Moon orbit and its
interpretation to be equivalent to the intrinsic gravitomagnetic
effects generated by the angular momentum of a central body
\cite{nor03}, are based and have been carried out by writing the
Moon's equation of motion in a frame comoving with the Sun (whereas
the orbit of the Moon is measured in a frame comoving with Earth,
i.e., the 'observable' quantity is the round-trip travel time of
laser pulses from Earth to Moon measured on Earth). To elucidate
with a simple example that the interpretation of term (2) depends on
the velocity of the frame of reference where the calculations are
performed, let us for example consider a single mass only, e.g., $M
\equiv M_\oplus$, i.e., the mass of Earth. In the weak-field and
slow-motion approximation of General Relativity, the corrections to
the Newtonian gravitational theory of order ${v \over c}^2 \sim {G M
\over {c^2 r}} << 1$ are described by the so-called 'post-Newtonian'
metric \cite{wil}. Let us then consider the post-Newtonian
expression of the Schwarzschild metric generated by $M_\oplus$. This
post-Newtonian metric can simply be obtained by expanding the
Schwarzschild metric at the lowest post-Newtonian order in $2 G M
\over {c^2 r}$. Let us then perform a local Lorentz transformation
with velocity $v^k$, in the new local frame we have the non-zero
gravitomagnetic metric components: $g_{0k} \sim {(G M_\oplus v^k)
\over {c^3 r}}$. In this new frame moving with velocity $v^k$ with
respect to $M_\oplus$, using the geodesic equation of motion of a
test-particle, e.g., the Moon, we then find the term $\sim \; {1
\over c^2} \vec{v}_{M} \times (\vec{v} \times \vec{G}_{M \oplus})$,
i.e., the frame-dependent term (2) of the Moon's equation of motion,
that is different from zero in this moving frame and is, however,
zero in a frame at rest relative to $M_\oplus$. This example shows
that the gravitomagnetic acceleration discussed in \cite{nor03,mnt},
i.e., the term (2), is a frame-dependent effect: when we go back to
the original frame, where the mass $M_\oplus$ is at rest, this term
of the Moon acceleration is zero.

Let us now consider the post-Newtonian metric generated by the
masses of both Sun and Earth, the non-diagonal components, $g_{0k}$,
of this metric, i.e., the components of the gravitomagnetic 'vector'
potential, are \cite{wil}: $g_{0k} = {7 \over 2} {G M_\oplus
v^k_\oplus \over {c^3 r_\oplus}}+ {1 \over 2} {G M_\oplus r^k_\oplus
({\vec r_\oplus \cdot \vec v_\oplus}) \over {c^3 r^3_\oplus}} + {7
\over 2} {G M_\odot v^k_\odot \over {c^3 r_\odot}}+ {1 \over 2} {G
M_\odot r^k_\odot ({\vec r_\odot \cdot \vec v_\odot}) \over {c^3
r^3_\odot}}$. In a frame of reference comoving with the Sun, the
components of the gravitomagnetic 'vector' potential are then:
$g_{0k} = {7 \over 2} {G M_\oplus v^k_\oplus \over {c^3 r_\oplus}}+
{1 \over 2} {G M_\oplus r^k_\oplus ({\vec r_\oplus \cdot \vec
v_\oplus}) \over {c^3
 r^3_\oplus}}$ and by using this gravitomagnetic potential in the
Moon's geodesic equation of motion, we find the acceleration (3),
i.e., the gravitomagnetic acceleration (2) of the Moon written in a
frame comoving with the Sun.

Nevertheless, the interpretation of the nature and meaning of term
(2) is simple in a frame comoving with Earth, where its measurable
effect is equivalent to the geodetic precession. Indeed, in a
geocentric frame of reference the gravitomagnetic 'vector' potential
components are: $g_{0k} = {7 \over 2} {G M_\odot v^k_\odot \over
{c^3 r_\odot}}+ {1 \over 2} {G M_\odot r^k_\odot ({\vec r_\odot
\cdot \vec v_\odot}) \over {c^3 r^3_\odot}}$ and by using this
gravitomagnetic potential in the Moon's geodesic equation of motion,
we then find the Moon's gravitomagnetic acceleration:

\begin{displaymath}
{4 \over c^2} \vec{v}^{(E)}_{M} \times (\vec{v}^{(E)}_\odot \times
\vec{G}_{M \odot}) \cong {4 \over c^2} \vec{v}^{(E)}_{M} \times
[\vec{v}^{(S)}_\oplus \times {G M_\odot \over R^3} (\vec{R} -
\vec{r}_{\oplus M})] =
\end{displaymath}

\begin{equation}
={4 \over c^2} [{G M_\odot \over R^3} (\vec{R} \times
\vec{v}^{(S)}_\oplus)] \times \vec{v}^{(E)}_{M} - {4 \over c^2}
\vec{v}^{(E)}_{M} \times (\vec{v}^{(S)}_\oplus \times {G M_\odot
\over R^3} \vec{r}_{\oplus M}) \;\;\;
\end{equation}

this acceleration of the Moon corresponds to the acceleration (2)
but it is written in a frame comoving with Earth: the first term of
the last expression of (6) is clearly equivalent to the geodetic
precession (5), apart from a numerical factor, and the second term
is today too small to be measured; of course, in order to describe
the Moon orbit, one must also include all the other terms in the
Moon's equation of motion, however, here we have only been
interested in analyzing the nature of the term (2) discussed in
\cite{nor03}.

A second argument shows that the interpretation of the term (2) as
an intrinsic gravitomagnetic effect is in fact frame-dependent. On
the other hand, in the case of a spacetime geometry generated by
both the angular momentum and the mass of a body (e.g., the Kerr
metric), the gravitomagnetic effects generated by the angular
momentum, for example that of Earth, are intrinsic to the spacetime
geometry and cannot be eliminated by a coordinate transformation,
indeed the angular momentum generates spacetime curvature, as
manifestly displayed by the curvature invariants.

This can rigorously be shown by using the curvature invariant ${\bf
^\ast R \cdot R}$, described in section 3, formally similar to the
electromagnetism invariant ${\bf  ^\ast F \cdot F} = {\vec E} \cdot
{\vec B}$. In the case of a point-mass metric generated by Earth and
Sun (neglecting the angular momenta of Earth and Sun that produce
effects that are presently unmeasurable on the Moon orbit) this
invariant is: $\sim \vec G \cdot \vec H$, where $\vec G$ is the
standard Newtonian electric-like field of Sun and Earth and $\vec H$
is the magnetic-like gravitational field; for a test-particle moving
with velocity $\vec v$, similarly to electrodynamics, this
magnetic-like gravitational field is $\sim {\vec v} \times {\vec G}$
and then, for any motion on the ecliptic plane, the invariant ${\bf
^\ast R \cdot R}$ is zero. Indeed, its expression, as calculated
\cite{math} in quasi-cartesian coordinates in a frame with origin at
the Sun ($x$ and $y$, and $z$ are respectively the coordinates on
and off the ecliptic plane), is at the lowest order in $G/c^2$:

\begin{equation}
{\bf ^\ast R \cdot R} \cong 288 {G^2 M_\oplus M_\odot \over {c^5
r^4_{M} \, r^4_{M \oplus}}} z_{M} (v^x_\oplus y_\oplus - v^y_\oplus
x_\oplus) (\hat{x}_{M} \cdot \hat{r}_{M \oplus})
\end{equation}

where $\vec x_\oplus$ and $\vec x_{M}$ are the position vectors of
Earth and Moon from the Sun and $z_{M}$ is the distance of the Moon
from the ecliptic plane; this expression is then: ${\bf ^\ast R
\cdot R} = 0$ on the ecliptic plane (even by considering that the
Moon orbit is slightly inclined of 5 degrees on the ecliptic plane,
its z component would only give a contribution to the change of its
radial distance from Earth of less than 1 $\%$ of the total change
discussed in \cite{nor03}).

\section{Acknowledgments}
I gratefully acknowledge the support of the Italian Space Agency.

\newpage

\pagestyle{empty}

\begin{figure}
\begin{center}
\includegraphics[scale=.8]{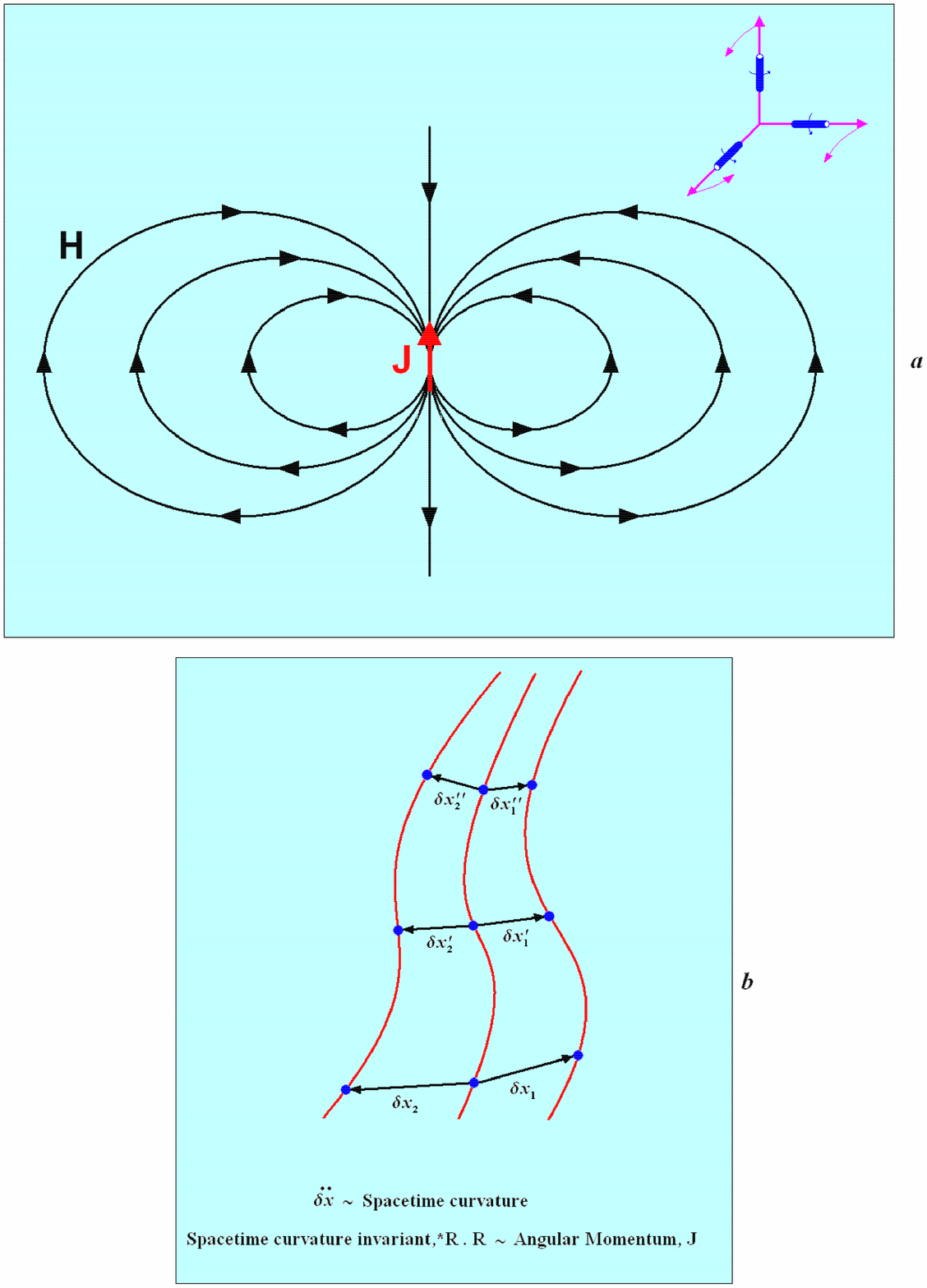}
\end{center}
\end{figure}

\newpage

{\bf Fig. 1} In panel $a$, I show the gravitomagnetic field
\cite{sch} ${\vec{H}}$ generated by the spin ${\vec{J}}$ of a
central body and the dragging of an inertial frame of reference
whose axes are determined by test-gyroscopes. In panel $b$, I show
how to get a measurement of the spacetime curvature and of the
gravitomagnetic field by local measurements only. One first measures
the relative accelerations, $\ddot{\delta x^{\alpha}}$, between a
number of test-particles which follow spacetime geodesics then,
using the geodesic deviation equation \cite{mtw,wei,ciuw}, one
obtains the spacetime curvature, i.e., one determines all the
components of the Riemann tensor, $R_{\alpha \beta \mu \nu}$.
Finally, using the Riemann tensor components, one obtains the
spacetime invariant discussed in the text that is a function of the
angular momentum and, in general, of the mass-energy currents. In
electromagnetism, using the Lorentz force equation, in order to
measure the six independent components of the electromagnetic field
tensor $F^{\alpha \beta}$, one needs to use at least two
test-particles endowed with electric charge \cite{mtw}. In General
Relativity, in order to measure the twenty independent components of
the spacetime curvature, i.e., of the Riemann tensor, the minimum
number of test-particles to be used is six, however, in vacuum, in
order to measure the ten independent components of the spacetime
curvature, is sufficient to use four test-particles
\cite{ciu86,ciud}.


\begin{thebibliography}{10}
\bibitem{math} The curvature invariants have been calculated using
MathTensor, a system for doing tensor analysys by computer  by L
Parker and S M  Christensen
\end{thebibliography}
\end{document}